\documentclass[aps,11pt,prd,showpacs,showkeys,preprintnumbers,superscriptaddress,nobibnotes,floatfix,longbibliography,notitlepage,nofootinbib]{revtex4-1}
\pdfoutput=1

\usepackage{afterpage}
\usepackage{placeins}
\usepackage{amsmath}
\usepackage{amsfonts}
\usepackage{bm}
\usepackage{amssymb}
\usepackage{mathrsfs}
\usepackage{cancel}
\usepackage{accents}
\usepackage{mciteplus,slashed}
\usepackage{amssymb,cancel,amsmath,relsize}
\usepackage{mathrsfs} 
\usepackage{dcolumn}
\usepackage{bm}
\usepackage[caption=false]{subfig} 
\usepackage{appendix}
\usepackage{feynmp-auto}
\usepackage[T1]{fontenc}	
\usepackage{csvsimple}
\usepackage{hyperref}
\usepackage[capitalise]{cleveref}
\usepackage{booktabs}
\usepackage{graphicx}
\usepackage{mathrsfs}
\usepackage{floatpag}
\usepackage[utf8]{inputenc}
\usepackage[dvipsnames]{xcolor}
\hypersetup{
    colorlinks,
    linkcolor={red!50!black},
    citecolor={blue!50!black},
    urlcolor={blue!80!black}
}
\usepackage[normalem]{ulem}
\usepackage{cleveref}
\usepackage{fancybox}
\usepackage{array}
\newcolumntype{P}[1]{>{\centering\arraybackslash}p{#1}}
\newcolumntype{M}[1]{>{\centering\arraybackslash}m{#1}}
\usepackage{tabularray}
\usepackage{titlesec}
\usepackage{multirow}

\newcommand {\ignore}[1]{}

\newcommand{\bmt}{\begin{pmatrix}}
\newcommand{\emt}{\end{pmatrix}}
\newcommand{\ba}{\begin{array}{c}}
\newcommand{\ea}{\end{array}}
\newcommand{\be}{\begin{equation}}
\newcommand{\ee}{\end{equation}}
\newcommand{\bea}{\begin{eqnarray}}
\newcommand{\eea}{\end{eqnarray}}

\newcommand{\bi}{\begin{itemize}}
\newcommand{\ei}{\end{itemize}}

\newcommand{\baz}{\begin{array}{cc}}
\newcommand{\besub}{\begin{subequations}}
\newcommand{\eesub}{\end{subequations}}

\def\e{\epsilon}

\def\q2 {q^2}

\def\bt{\begin{table}}
\def\et{\end{table}}

\titleformat*{\paragraph}{\bfseries \itshape}
\titleformat*{\section}{\centering\bfseries }
\titleformat*{\subsection}{\centering\bfseries }

\graphicspath{{./figures/}}

\begin{document}
\title{\Large \textbf{Constraining Gravitational Waves from Superconducting Cosmic Strings with the Dark Ages $21$-cm Signal}}

\author{Adeela Afzal}
\affiliation{Bogoliubov Laboratory of Theoretical Physics, Joint Institute for Nuclear Research\\ 141 980, Dubna, Moscow Region, Russia}
\email{adeelaafzal@theor.jinr.ru}

\begin{abstract}
Superconducting cosmic strings (SCSs) are characterized by three parameters: the dimensionless string tension $G\mu$, the current amplitude $Y$, and the vector coupling, $\tilde{e}$ as recently explored in \cite{Rybak:2024our}. Using the Charge-Velocity-dependent One-Scale (CVOS) model, we compute the stochastic gravitational wave background (SGWB) emitted by current-carrying loops, including the suppression effect of vector radiation. While previous studies have shown that large $\tilde{e}$ suppresses the SGWB, allowing larger $G\mu$ to evade pulsar timing array (PTA) bounds \cite{Rybak:2024our}, we demonstrate that this parameter space is definitively constrained by an independent, astrophysically clean probe: the Dark Ages global $21$-cm signal. We recalculated the recently derived $21$-cm bounds \cite{Si:2025vsj} into the generic $(\tilde{e},Y,G\mu)$ parameter space and find that at the threshold where vector radiation emission starts dominating, any coupling $\tilde{e}\gtrsim 10^{-7}$ and $Y\simeq 0.67$ and $G\mu\gtrsim 10^{-13}$ injects sufficient ionizing radiation to completely erase the $21$-cm absorption signal at redshift $z\simeq89$. Greater current amplitudes impose more stringent constraints. This excludes a large parameter space considered viable. Our model-independent results highlight the complementarity between future lunar-based $21$-cm experiments and gravitational wave (GW) observatories.
\end{abstract}

\maketitle


\section{Introduction}
\label{sec:introduction}
Cosmic strings (CSs) are one-dimensional topological defects that arise in a wide variety of particle physics models, including grand unified theories (GUTs) and their gauge $U(1)$ extensions. 
Recent treatments of topological structures within $E_6$ and $SO(10)$ can be found in Refs.~\cite{KIBBLE1982237,Lazarides:2019xai,Lazarides:2023iim,Afzal:2023kqs,Maji:2024pll,Maji:2025thf,Dunsky:2021tih,Martin:1996ea, Lazarides:2022jgr, Maji:2025itv} and a number of realistic models, primarily based on $SO(10)$ and its various subgroups, have been proposed to account for the SGWB, see Refs.~\cite{Buchmuller:2020lbh,Buchmuller:2021mbb,Auclair:2022ylu,Afzal:2022vjx,Ahmed:2022rwy,Saad:2022mzu,Lazarides:2022spe,Lazarides:2022ezc,DiBari:2023mwu,Buchmuller:2023aus,Antusch:2023zjk,Lazarides:2023rqf,Ahmed:2023rky,Maji:2023fhv,Ahmed:2023pjl,Afzal:2023cyp,Fu:2023mdu,Lazarides:2023ksx,Lazarides:2023bjd,Maji:2024cwv,Roshan:2024qnv,Afzal:2025lpx,Tranchedone:2026lav}.
A cosmic string is said to be superconducting when it carries a persistent electric current, typically arising from the existence of charged fermion or scalar zero modes propagating along the string core~\cite{Witten:1984eb,Maji:2025itv}. Such currents can be generated either through the capture of cosmic charges or via the Kibble mechanism \cite{T_W_B_Kibble_1976} during the string-forming phase transition. The presence of this current significantly alters the string's equation of state, modifies its dynamical evolution, and opens up additional energy-loss channels through the emission of vector (electromagnetic) radiation. These objects are therefore characterized by two additional parameters beyond the standard string tension $G\mu$: the current amplitude $Y$, and the dimensionless vector coupling $\tilde{e}$, which governs the efficiency of electromagnetic radiation \cite{Rybak:2024our}. The cosmological evolution and observational signatures of superconducting CSs (SCSs) can be studied in a model-independent way using the Charge-Velocity-dependent One-Scale (CVOS) model ~\cite{Martins:2020jbq,Martins:2021cid,Rybak:2023jjn}.

SCS loops may emit both gravitational waves (GWs) and vector (electromagnetic) radiation. The efficiency of vector radiation, governed by $\tilde{e}$, can significantly suppress the GW spectrum~\cite{Rybak:2024our}. This suppression opens a parameter window where the stable strings with relatively large tension (e.g., $G\mu\lesssim 10^{-8}$) evade current PTA bounds, such as those from NANOGrav \cite{NANOGrav:2023hvm} while remaining consistent with the CMB~\cite{Lasky:2015lej} and the JWST~\cite{Blamart:2025szc} constraints.
In this work, we show that this large parameter space considered to be viable is strongly constrained by an independent probe: the Dark Ages global $21$-cm signal. During the Dark Ages ($z\sim40-200$), the intergalactic medium is free from astrophysical uncertainties, making it a pristine laboratory for exotic energy injection. Ionizing radiation emitted by SCS loops heats and ionizes the gas, suppressing the $21$-cm absorption signal \cite{Cyr:2023iwu, Theriault:2021mrq, Cyr:2023yvj}. We have recalculated the recently derived $21$-cm bounds~\cite{Si:2025vsj} into the $(\tilde{e},Y,G\mu)$ space, we demonstrate that the parameter values required to evade PTA for large $G\mu$ values are firmly constrained. Our results reveals that any coupling $\tilde{e}\gtrsim 10^{-7}$ for $Y\simeq 0.67$ and $G\mu\gtrsim 10^{-13}$ injects sufficient ionizing radiation to completely erase the $21$-cm absorption signal at redshift $z\simeq89$ and greater current amplitudes impose more stringent constraints as presented in Fig.~\ref{fig:21cmsignalbounds}. Our model-independent analysis demonstrates the synergy between future lunar-based $21$-cm experiments~\cite{Rapetti:2019lmf, burns2020transformative} and GW detectors.

In this paper we first explore in Sec.~\ref{sec:CCCS} the GW spectrum generated by SCSs, taking into account any chiral currents associated with them following Ref.~\cite{Rybak:2024our}. In Sec.~\ref{sec:21cm}, we present the Dark Ages 21-cm signal constraints on SCSs. Finally, we conclude in Section \ref{sec:conclusion}.

\section{Superconducting Cosmic Strings and Gravitational Wave Production}
 \label{sec:CCCS}
 We consider the strings carrying the chiral current based on the  Nambu-Goto action, the CSs are infinitely thin and are coupled to a gauge vector field, as proposed in Refs.~\cite{WITTEN1985557,PhysRevD.45.1091}. To describe the evolution of the CS network, we adopt the CVOS model. For a detailed discussion, we refer the reader to Refs.~\cite{Martins:2020jbq, Martins:2021cid, Rybak:2023jjn, Rybak:2024our}. We also assume that the major contribution to the SGWB arises from the current-carrying loops characterized by the loop size parameter, $\alpha=0.1$, with quasi-cusps \cite{Rybak:2024our}. With superconductivity, the cusp-like feature is replaced by a quasi-cusp, where the string still attains large Lorentz factors but does not strictly reach the speed of light Ref.~\cite{Rybak:2024our}. We sum over $150$ harmonic modes while computing the GW spectra since the higher harmonics provide a negligible contribution (for details we refer the reader to Fig. 17 of Ref.~\cite{Rybak:2024our}).
 
\begin{figure}[t]
    \centering
     \includegraphics[width=0.97\textwidth]{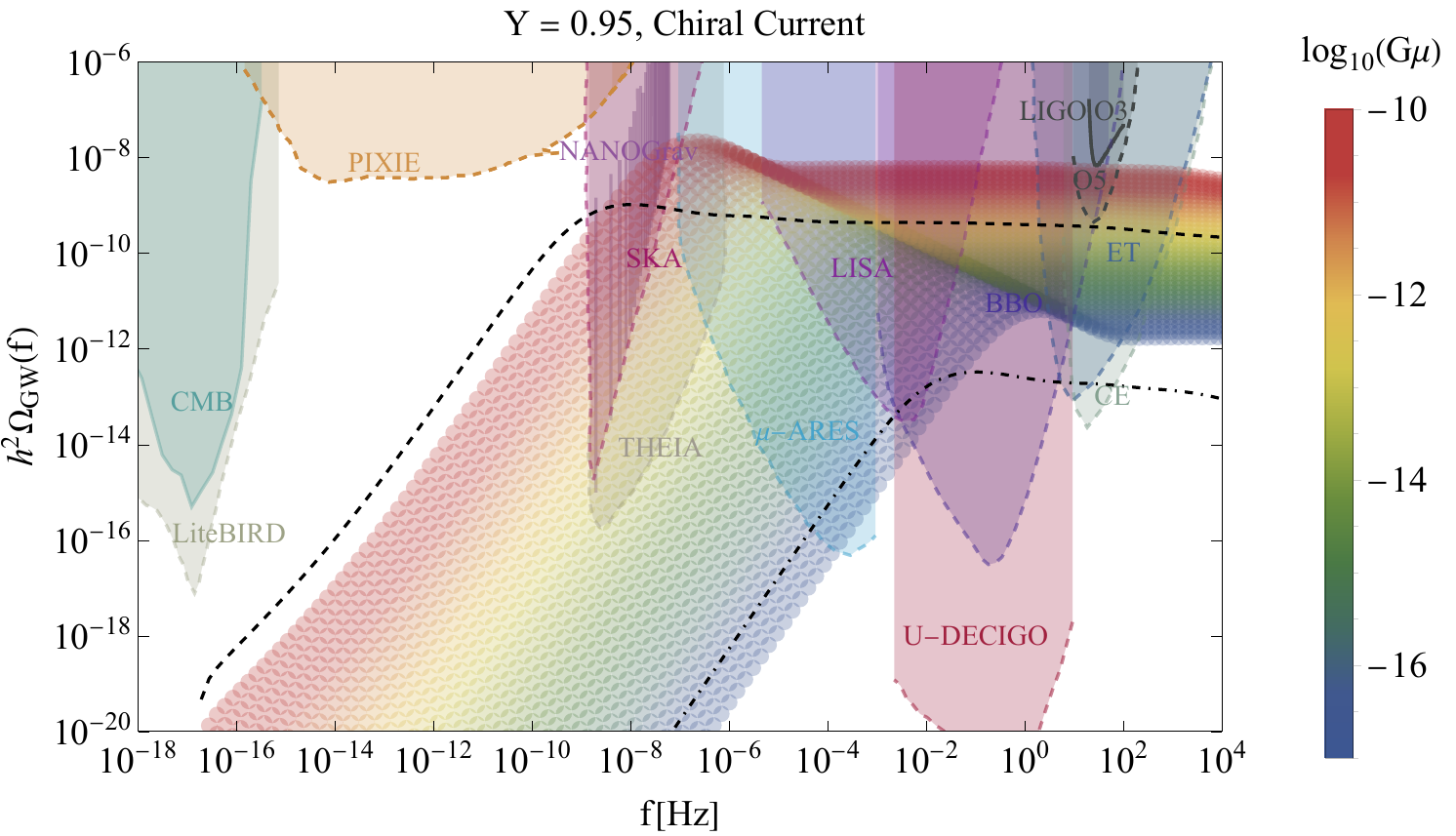}
     \caption{GW spectra of stable chiral current CSs for $G\mu\simeq 10^{-17}$ to $10^{-10}$, shown by the vertical colored bar for $Y\simeq 0.95$. The colored shaded regions indicate the sensitivity curves of the present (solid boundaries) LIGO O3 \cite{KAGRA:2021kbb}, NANOGrav \cite{NANOGrav:2023gor}, CMB \cite{Lasky:2015lej} and future (dashed boundaries) LiteBIRD \cite{LiteBIRD:2022cnt}, PIXIE \cite{A_Kogut_2011}, SKA \cite{Smits:2008cf}, THEIA \cite{Garcia-Bellido:2021zgu}, $\mu$-ARES \cite{Sesana:2019vho}, LISA \cite{amaroseoane2017laser}, BBO \cite{Corbin:2005ny}, U-DECIGO \cite{Yagi:2011wg, Kawamura:2020pcg}, ET \cite{Punturo:2010zz}, CE \cite{Reitze:2019iox} experiments. The standard Nambo-Goto strings for $G\mu\simeq10^{-10}\,(10^{-17})$ are displayed with black dashed (dot-dashed) as a reference.}
     \label{fig:StableGmu10to17Yrd0p95etild0}
 \end{figure}

Following~\cite{Rybak:2024our}, the current associated with the CSs is quantified by the charge amplitude,
\begin{align}
    \mathcal{Y}=\dfrac{Y}{2}\left[1- \text{Tanh}\left(\dfrac{\text{log}(a)+7(1-Y}{3}\right)\right],
\end{align}
 with $a$ being the scale factor, and $0\leq Y \lesssim 1 $, is the value of the current amplitude during the radiation-dominated epoch. It was previously explored in Refs.~\cite{Rybak:2022sbo,Afzal:2023kqs} (for stable and metastable strings, respectively) that the loops carrying a certain amount of current may enhance the amplitude of the GW spectrum due to the non-trivial evolution of the VOS variables. We have shown this effect for a range of $G\mu=10^{-17}$ to $10^{-10}$ in Fig.~\ref{fig:StableGmu10to17Yrd0p95etild0} for a fixed $Y\simeq 0.95$ in order to be consistent with the LIGO O3 run \cite{KAGRA:2021kbb} for larger $G\mu$. As a consequence of the substantial current, even a relatively small value of string tension such as $G\mu\simeq 10^{-17}$ falls within the readily observable range of GW detectors. This is not the case for the standard Nambo-Goto strings, as shown by the black dashed (dot-dashed) line for $G\mu\simeq 10^{-10}\,(10^{-17})$.

 Next we consider the strings that are coupled to a gauge vector field that may contribute to the emission of vector radiation, as recently explored in Ref.~\cite{Rybak:2024our}. Following their analysis, the emission of vector radiation by superconducting CS loops can be characterized by the dimensionless efficiency parameter $\Gamma^{\rm em}$, defined as the total power radiated in vector fields
\begin{align}
    P^{\rm em}=e^2 \Gamma^{\rm em},
\end{align}
where $e$ is the dimensionful coupling between the current carriers and
the vector fields. It is convenient to define a dimensionless coupling constant $\tilde{e}^2\equiv e^2/\mu$, with $\mu$ being the string tension. The inclusion of vector radiation fundamentally alters the decay of superconducting CS loops. The total power emitted by loops is defined as \cite{Rybak:2024our}
 \begin{align}
 \label{eq:powerloop}
     P(\mathcal{Y})\equiv G\mu\,\Gamma^{\rm gr}(\mathcal{Y}) + \tilde{e}^2\,\Gamma^{\rm em}(\mathcal{Y}).
 \end{align}
\begin{figure}[t]
     \centering    \includegraphics[width=0.9\textwidth]{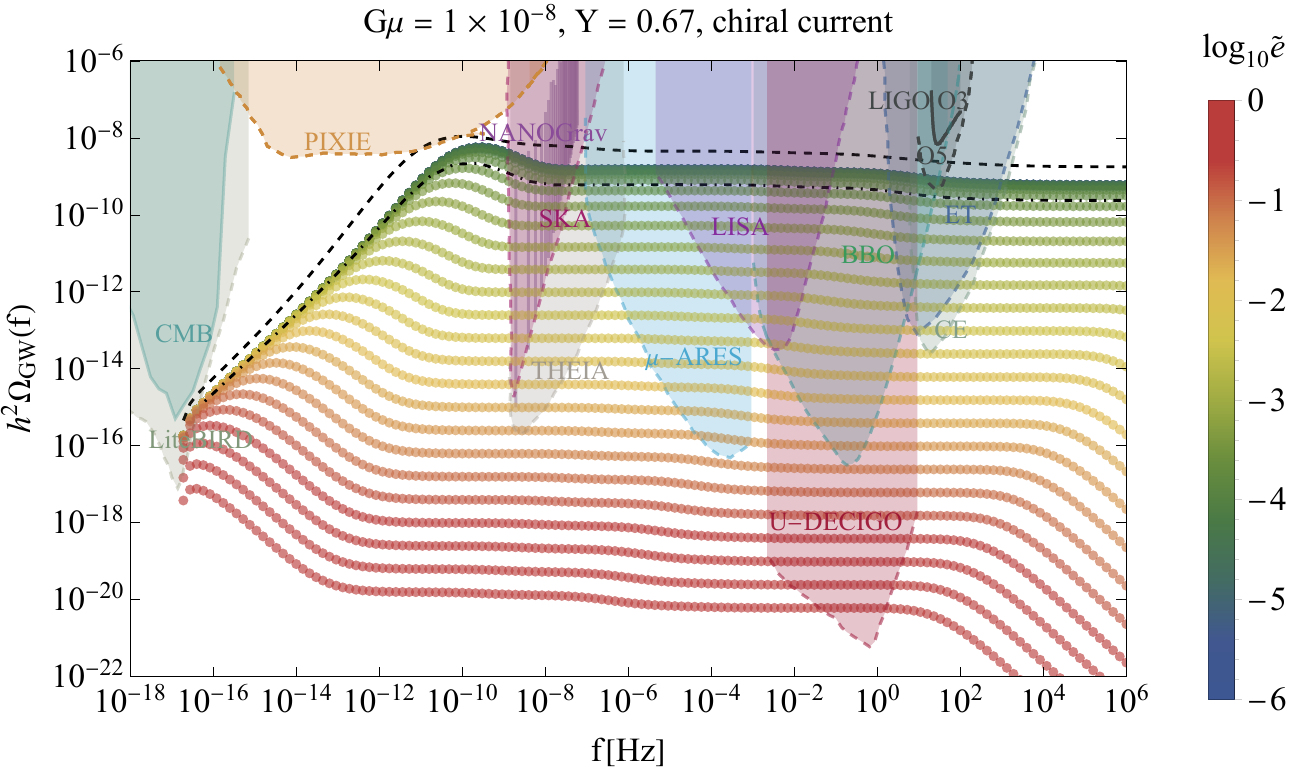}
     \caption{GW spectra of stable superconducting CSs for $G\mu\simeq 10^{-8}$ taking account of non-zero $\tilde{e}$, the charge of the current carriers. These CSs carry the chiral current, and the associated emission of vector radiation is shown by the vertical colored bar denoted by $\tilde{e}$, for $Y=0.67$. The black dot-dashed line corresponds to $\tilde{e}\simeq 2\times10^{-4}$ and at and below this line, the displayed parameter space is not in conflict with the bounds from NANOGrav~\cite{NANOGrav:2023hvm}. The standard Nambu-Goto string network case with $Y=0=\tilde{e}$ is included with a black dashed line for comparison. The colored shaded regions indicate the sensitivity curves of the present (solid boundaries) LIGO O3~\cite{KAGRA:2021kbb}, NANOGrav~\cite{NANOGrav:2023gor}, CMB~\cite{Lasky:2015lej} and future (dashed boundaries) LiteBIRD~\cite{LiteBIRD:2022cnt}, PIXIE~\cite{A_Kogut_2011}, SKA~\cite{Smits:2008cf}, THEIA~\cite{Garcia-Bellido:2021zgu}, $\mu$-ARES~\cite{Sesana:2019vho}, LISA~\cite{amaroseoane2017laser}, BBO~\cite{Corbin:2005ny}, U-DECIGO~\cite{Yagi:2011wg,Kawamura:2020pcg}, ET~\cite{Punturo:2010zz}, CE~\cite{Reitze:2019iox} experiments.}
    \label{fig:StableGmu8_y0p6_evar_chiral}
\end{figure}
 Here, $\Gamma^{\rm gr}(\mathcal{Y}) = \Gamma^{\rm gr}_0(1-\sqrt{\mathcal{Y}})^B$ and $\Gamma^{\rm em}(\mathcal{Y}) = \Gamma^{\rm em}_0\sqrt{\mathcal{Y}}\,(1-\sqrt{\mathcal{Y}})^D$ are, respectively, the efficiency of GW and vector radiation emission. 
  This additional contribution will cause a suppression of the GW spectrum. The numerical values for these efficiency parameters are: $\Gamma^{\rm gr}_0\simeq 50$, $\Gamma^{\rm em}_0=8.6$, $D=1.1$ and $B=\sqrt{2}$ for chiral loops with quasi-cusps \cite{Rybak:2024our}. The dependence of the spectral energy density of gravitational radiation on the total power emitted by the loops is \cite{Rybak:2024our},
 \begin{align}
     \Omega_{\rm GW}\propto \left(\alpha\dfrac{d\xi}{dt}+P(\mathcal{Y})\right)^{-1},
 \end{align}
 where $\xi$ is the characteristic length of the network and $t$ is the time of loop creation. It is worth mentioning that for  smaller values of $G\mu$, the emission of vector radiation becomes a dominant energy loss mechanism. However, for larger loops ($\alpha=0.1$), both $G\mu\,\Gamma^{\rm gr}(\mathcal{Y})$ and $\tilde{e}^2\,\Gamma^{\rm em}(\mathcal{Y})$ should be taken into account.

 In Fig.~\ref{fig:StableGmu8_y0p6_evar_chiral}, we present the GW spectra of stable CSs for $G\mu\simeq 10^{-8}$ with the inclusion of both $Y$ and $\tilde{e}$, and the standard case of Nambu-Goto. $Y=0=\tilde{e}$ (black dashed line) has also been included for comparison. We have set the value of $Y=0.67$ for which the amplitude is minimum due to the associated current to lie within the sensitivity reach of current and future planned experiments as presented by colored shaded regions with dashed and solid lines in Fig.~\ref{fig:StableGmu8_y0p6_evar_chiral}. The interval $ 0 < Y < 0.67$ is the decreasing and $0.67 < Y < 1$ increasing interval for the GW spectrum amplitude; we refer the reader to Fig.~8 (3) of Ref.~\cite{Rybak:2022sbo}(\cite{Rybak:2024our}) and the discussion therein. The trend in Fig.~\ref{fig:StableGmu8_y0p6_evar_chiral} indicates a trade-off between the emission of GWs and vector radiation, with stronger vector radiation corresponding to the weaker GWs. Note that standard stable CSs with tension $G\mu\gtrsim10^{-10}$ are ruled out by NANOGrav 15-year data, since their predicted GW spectra exceed the observational upper limit derived from the non-observation of correlated timing residuals in the pulsar timing arrays data \cite{NANOGrav:2023hvm}. This demonstrates that if the strings are sufficiently superconducting ($\tilde{e}\gtrsim 2\times 10^{-4}$), the upper limit on the string tension is $G\mu\lesssim 10^{-8}$ remain viable (from CMB \cite{Lasky:2015lej, Raidal:2026cpb,Caloni:2026dyu} and NANOGrav \cite{NANOGrav:2023gor}) and the spectrum lies within the observable range. But in the Sec.~\ref{sec:21cm} we demonstrate that this parameter space is constrain by the Dark Ages global $21$-cm signal.
\section{Dark Ages $21$-cm Constraints}
\label{sec:21cm}
In this section, we investigate the impact of the radiating SCSs on the global $21$-cm absorption signal and obtain constraints on ($\tilde{e}$ and $G\mu$) for a fixed value of $Y$. The $21$-cm line arises from the hyperfine transition in neutral hydrogen, corresponding to a rest frame frequency of $1420$~MHz. The differential brightness temperature of the $21$-cm signal against the cosmic microwave background (CMB) is given by \cite{Pritchard_2012,DAmico:2018sxd,Mitridate:2018iag,Cyr:2023iwu,Nishizawa:2024bnh,Si:2025vsj}
\begin{align}
    \delta T_{21}\simeq 27 N_{\text{HI}}\left(1-\dfrac{T_R}{T_s}\right)\left(\dfrac{0.15(1+z)}{10\,\Omega_m}\right)^{0.5}\left(\dfrac{h\,\Omega_b}{0.023}\right)\text{mK},
\end{align}
where $N_{\text{HI}}$ is the neutral hydrogen fraction, $T_R$ is the background radiation temperature typically set
by the CMB, and $T_s$ is the spin temperature. We follow the standard $\Lambda$CDM framework and use $\Omega_b=0.04859$, $\Omega_m=0.315$, $\Omega_r=10^{-4}$, $h=0.68$ and $H_0=100 \times h$ \cite{Planck:2018vyg}. The spin temperature is determined by the balance between collisional coupling, Wouthuysen-Field coupling, and coupling to the CMB given by~\cite{Pritchard_2012}
\begin{align}   T_s^{-1}=\dfrac{T_R^{-1}+x_c\,T_\text{gas}^{-1}+x_\alpha\,T_\alpha^{-1}}{1+x_c+x_\alpha}.
\end{align}
Here $T_\text{gas}$ is the intergalactic gas (IG) kinetic temperature, $T_\alpha$ is the Lyman-$\alpha$ color temperature, $x_c$ and $x_\alpha$ are the collisional and Wouthuysen-Field coupling coefficients, respectively \cite{4065250,1952AJ.....57R..31W}. During the Dark Ages ($40\lesssim z\lesssim200$), the universe is largely homogeneous and free from astrophysical sources. The IG cools adiabatically as $T_\text{gas}\propto(1+z)^2$, while the CMB cools as $T_\gamma\propto(1+z)$. Collisional coupling keeps the spin temperature coupled to the gas until $z\sim40$, producing an absorption signal with amplitude $\delta T_{21}\sim-42$~mK at $z\sim89$ in the standard $\Lambda$CDM framework \cite{Mohapatra:2024djd}. The absence of astrophysical uncertainties during this epoch makes the Dark Ages $21$-cm signal an exceptionally clean probe of exotic energy injection, including that from SCSs \cite{Si:2025vsj}.
\subsection{Energy Injection from Superconducting Cosmic Strings}
\label{subsec:energyinjection_SCSs}
SCS loops emit electromagnetic radiation across a broad spectrum. Let $\ell$ be the invariant loop length that decrease as a result of the emission of gravitational and vector radiation as follows \cite{Rybak:2024our}
\begin{align}
    \dot{\ell}= -G\mu\,\Gamma^{\rm gr}(\mathcal{Y}) - \tilde{e}^2\,\Gamma^{\rm em}(\mathcal{Y}).
\end{align}
The efficiency of GW $\Gamma^{\rm gr}(\mathcal{Y})$ and vector radiation $\Gamma^{\rm em}(\mathcal{Y})$ is defined after Eq.~\ref{eq:powerloop}. Following Refs.~\cite{Tashiro:2012nv}, the spectrum of photons with frequency $\omega$ emitted from a cusp on a loop of length $\ell$ in terms of new parameter, $\tilde{e}$, can be written as
\begin{align}
    \dot{N_\omega}\equiv \dfrac{d^2N}{d\omega dt}\simeq \dfrac{4\pi \tilde{e}^4\mu\,\ell^{1/3}}{3\,\omega^{5/3}}.
\end{align}
The volumetric energy injection rate in the matter-dominated era becomes \cite{Tashiro:2012nv,Si:2025vsj}
\begin{align}
    \dfrac{d^2E}{dVdt}\equiv \int_0^\omega d\omega\,\omega \int_0^\infty dN\,\dot{N_\omega} = A(t)\, g(Y,\tilde{e},G\mu)\,\omega^{1/3},
\end{align}
where $A(t) \propto t^{-19/6}$ encodes the time dependence, and the degenerate parameter
\begin{align}
\label{eqn:injeffparam}
    g(Y,\tilde{e},G\mu)\equiv \tilde{e}^4\,\mu \left(\Gamma_\text{tot}(\mathcal{Y}) G\mu\right)^{-7/6},
\end{align}
controls the overall energy injection efficiency. The total decay rate, including both gravitational and electromagnetic emission, is defined to be
\begin{align}
\label{eqn:gammatot}
    \Gamma_\text{tot}(\mathcal{Y})= \Gamma^{\rm gr}(\mathcal{Y})+\Gamma^{\rm em}(\mathcal{Y})\dfrac{\tilde{e}^2}{G\mu}.
\end{align}
We proceed with the case of chiral loops with quasi-cusps. The spectrum includes both radio and ionizing photons. Photons with energy less than $13.6$~eV cannot ionize hydrogen atoms in the ground state, however those in the $13.6$~eV to $10^4$~eV range deposit energy through heating. The energy available for ionization and heating is \cite{Si:2025vsj}
\begin{align}
\label{eq:dEasg}
     \dfrac{d^2E}{dVdt}=\mathcal{F}(\omega,z)A(t) g(Y,\tilde{e},G\mu)\{\omega^{1/3}_{10^4\,\text{eV}}-\omega^{1/3}_{13.6\,\text{eV}}\},
\end{align}
where $\mathcal{F}(\omega,z)$ is the deposition efficiency \cite{PhysRevD.87.123513,PhysRevD.93.023521}.
\subsection{The $21$-cm Constraint}
\begin{figure}[t]
    \centering
     \includegraphics[width=0.97\textwidth]{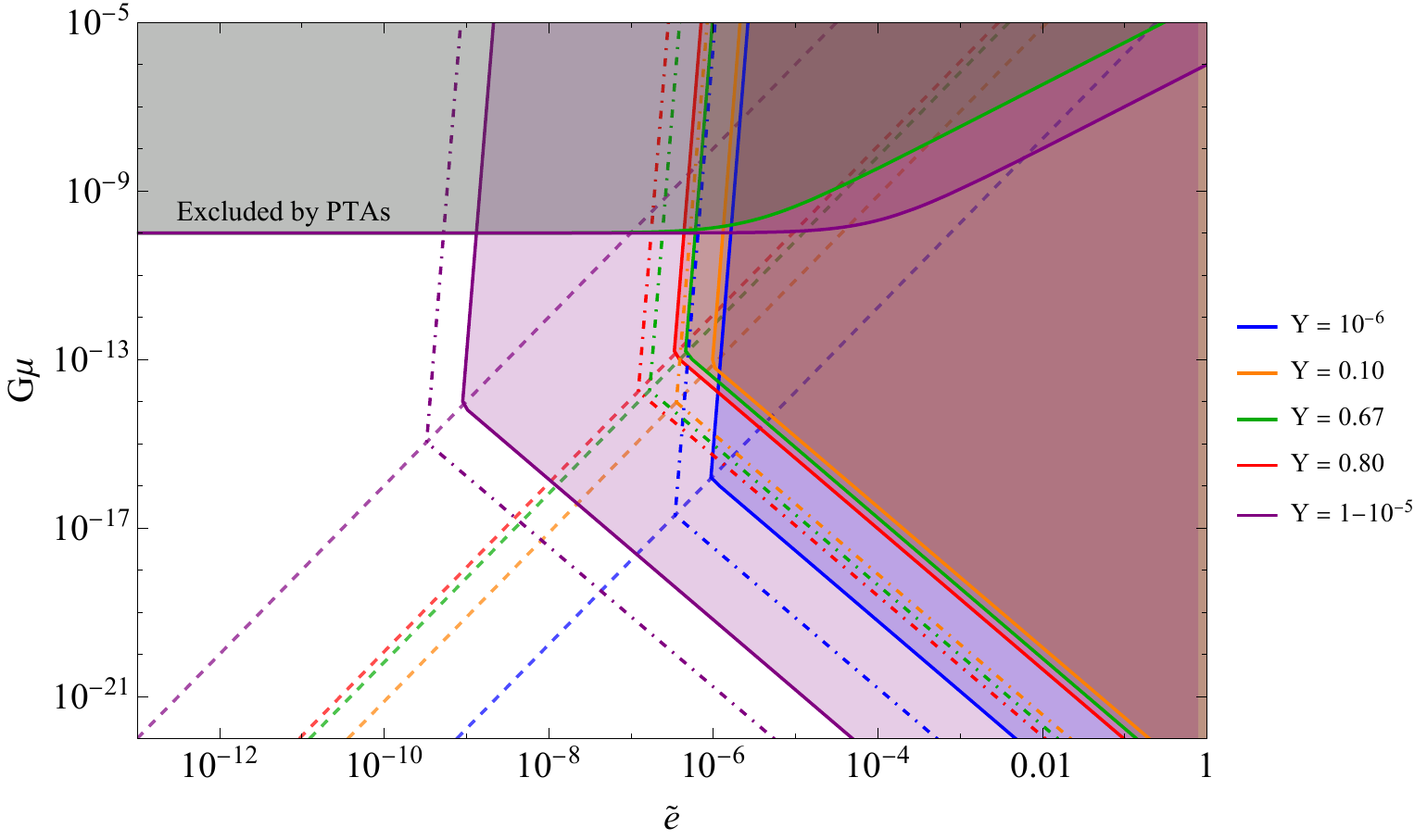}
     \caption{Combined constraints on SCSs in the ($\tilde{e}, G\mu$) plane for various values of the current amplitude $Y$, as indicated in the legend. The solid curves correspond to the Dark Ages $21$-cm constraint for a forecasted sensitivity of $\Delta T_{21}=5$~mK ($g\simeq 1.3 \times 10^{13}\,\text{GeV}^2$), while the dot-dashed curves correspond to $\Delta T_{21}=15$~mK ($g\simeq 5.1 \times 10^{14}\,\text{GeV}^2$). The shaded parameter space is excluded by the $21$-cm bound for the corresponding value of $Y$. The dashed lines indicate the critical coupling $\tilde{e}_c$ where gravitational and electromagnetic radiation efficiencies are equal. The blue shaded region denotes the excluded parameter space from the most conservative $21$-cm bound ($5$~mK sensitivity) for $Y=1-10^{-5}$, while the gray~($Y=0.67$), purple~($Y=1-10^{-5}$) shaded region indicates the PTA (NANOGrav \cite{NANOGrav:2023hvm}) exclusion for reference. The $21$-cm constraints are remarkably stringent: for $Y\simeq0.67$ and $G\mu\gtrsim 10^{-13}$, any coupling $\tilde{e}\gtrsim 10^{-7}$ is excluded; for larger currents ($Y\simeq 1-10^{-5}$), the bound tightens to $\tilde{e}\gtrsim 8 \times 10^{-10}$.}
     \label{fig:21cmsignalbounds}
 \end{figure}
The deposited energy modifies the thermal and ionization history of the IG through the coupled differential equations for $T_\text{gas}(z)$ and ionization fraction, $x_e(z)$ \cite{Natwariya:2022xlv,Nishizawa:2024bnh,PhysRevD.111.043002,Mitridate:2018iag,Ali_Ha_moud_2011,PhysRevD.92.083528,Si:2025vsj}
\begin{align}
\dfrac{dT_\text{gas}}{dz}
&= \left.\dfrac{dT_\text{gas}}{dz}\right|_{\text{std}}
- \left.\dfrac{2}{3(1+z)H(z)}\dfrac{1+2x_e}{3N_b^\text{tot}}\dfrac{d^2E}{dVdt}\right|_{\text{dep}},\\\notag
\dfrac{dx_e}{dz}&= \left.\dfrac{dx_\e}{dz}\right|_{\text{std}}
- \left.\dfrac{1-x_e}{(1+z)H(z)N_b^\text{tot}} \left(\dfrac{\mathcal{C}}{E_0}+\dfrac{1-\mathcal{C}}{E_\alpha}\right)\dfrac{d^2E}{dVdt}\right|_{\text{dep}},
\end{align}
where the standard (std) terms describe the usual recombination and cooling history~\cite{Si:2025vsj}. For a given $g(Y,\tilde{e},G\mu)$ (Eq.~\ref{eq:dEasg}), solving these equations yields $T_\text{gas}(z)$ and $x_e(z)$ from which the $21$-cm signal $\delta T_{21}(z)$ can be computed. 
The authors of Ref.~\cite{Si:2025vsj} have performed this analysis (we refer the readers to Fig.~3 of \cite{Si:2025vsj} for detail discussion) and found that the standard $\Lambda$CDM prediction at $z\sim 89$ is $\delta T_{21}\sim -42$~mK corresponding to $g\simeq 2.2 \times 10^{12}\text{GeV}^2$ at and below this value the effect of decaying SCS on $21$-cm
signal is no longer detectable. As the energy injection parameter increases,  $g\gtrsim 5.1 \times 10^{14}\text{GeV}^2$, 
the cosmic dawn $21$-cm signal becomes an emission signal, which is unphysical and thus excluded \cite{Si:2025vsj}. To place quantitative constraints on SCS parameters, the authors consider the forecasted sensitivities of future lunar-based $21$-cm experiments~\cite{burns2020transformative,Rapetti:2019lmf}. Based on the calculation in Ref.~\cite{Si:2025vsj}, the critical values of $g$ are
\begin{align}
    g = \begin{cases}
    1.3 \times 10^{13}\,\,\,\text{GeV}^2& \text{if } \Delta T_{21}=5\,\,\text{mK} \\
 5.1 \times 10^{14}\,\,\,\text{GeV}^2& \text{if } \Delta T_{21}=15\,\,\text{mK}
\end{cases}.
\end{align}
Values of $g$ exceeding these thresholds would produce deviations larger than the experimental uncertainty and are therefore excluded \cite{Si:2025vsj}.

There exist a critical value of $\tilde{e}_c$ at which the gravitational radiation is equal to electromagnetic radiation for a given value of $G\mu$. From Eq.~\ref{eqn:gammatot}, the critical value can be written as, 
\begin{align}
    \tilde{e}^2_c\equiv \dfrac{\Gamma^\text{gr}(\mathcal{Y})}{\Gamma^\text{em}(\mathcal{Y})}\,G\mu.
\end{align}
For values above this critical threshold, gravitational radiation dominates over the electromagnetic counterpart. This boundary is indicated by the dashed lines in Fig.~\ref{fig:21cmsignalbounds}, which correspond to the specific values of $Y$ given in the legend. The total decay rate Eq.~\ref{eqn:gammatot} can now be expressed as
\begin{align}
    \label{eqn:gammatotec}
    \Gamma_\text{tot}(\mathcal{Y})= \Gamma^{\rm gr}(\mathcal{Y})\left(1+\dfrac{\tilde{e}^2}{\tilde{e}_c^2}\right).
\end{align}
The energy injection efficiency parameter Eq.~\ref{eqn:injeffparam} can be rewritten as
\begin{align}
\label{eqn:injeffparammodify}
    g(Y,\tilde{e},G\mu)=\tilde{e}^4\,\mu \left(\Gamma^\text{gr}(\mathcal{Y}) \left(1+\dfrac{\tilde{e}^2}{\tilde{e}_c^2}\right)G\mu\right)^{-7/6}.
\end{align}
For two regime above and below the critical threshold, Eq.\ref{eqn:injeffparammodify} gives
\begin{align}
    \tilde{e}^2 = \begin{cases}
    (g\,G)^{1/2}\left(\Gamma^\text{gr}(\mathcal{Y})^7\,G\mu\right)^{1/12}& \text{if } \tilde{e}^2\ll \tilde{e}_c^2 \\
 (g\,G)^{6/5}\left(\dfrac{\Gamma^\text{gr}(\mathcal{Y})^{7}\,G\mu}{\tilde{e}_c^2}\right)^{1/5}& \text{if } \tilde{e}^2\gg \tilde{e}_c^2 
\end{cases}.
\end{align}
These constraints have been presented in Fig.~\ref{fig:21cmsignalbounds}.
The solid and dot-dashed curves correspond to $g\simeq 1.3 \times 10^{13}\,\,\text{GeV}^2$ and $g\simeq 5.1 \times 10^{14}\,\,\text{GeV}^2$ respectively, while the color coding indicates the various $Y$ values specified in the legend. For a small current amplitude, $Y\simeq 10^{-6}$, we find that the parameter region $\tilde{e}\lesssim 10^{-6}$ remains viable. As discussed in Sec.~\ref{sec:CCCS}, the GW amplitude is an increasing function of $Y$ over the interval $0.67<Y<1$. This trend is clearly reflected in Fig.~\ref{fig:21cmsignalbounds}: for larger values of $Y$, specifically $Y\simeq1-10^{-5}$, the bounds become considerably more stringent, ultimately restricting the allowed parameter space to $\tilde{e}\lesssim 8\times10^{-10}$. Consequently, the parameter space previously considered viable for evading PTA constraints is firmly excluded by the Dark Ages $21$-cm signal. This demonstrates a complementarity between future lunar-based $21$-cm experiments and GW observatories in constraining SCSs models.

Note that the limit $Y=1$ is a mathematical boundary corresponding to current saturation. In this limit, both $\Gamma^{\rm gr}$ and $\Gamma^{\rm em}$
  vanish, the loop oscillation period diverges, and the string effectively becomes a non-radiating vorton \cite{Rybak:2024our}. Dynamically, the network asymptotically approaches but never reaches exactly $Y=1$. Therefore, in our numerical analysis, we consider values arbitrarily close to unity, such as $Y=1-10^{-5}$, to approximate this saturation limit. For completeness, we note two limiting cases. First, in the limit $Y=0$, the string carries no current. From Eq.~\ref{eq:powerloop}, we have $\Gamma^\text{gr}(0) = \Gamma^\text{gr}_0$ and $\Gamma^\text{em}(0) = 0$, so the total power emitted by loops reduces to $P = \Gamma^\text{gr}_0\,G\mu$. The GW spectrum therefore reduces to the standard Nambu-Goto case, which is firmly ruled out by NANOGrav for $G\mu\gtrsim 10^{-10}$~\cite{NANOGrav:2023hvm}. Second, in the limit $\tilde{e}=0$, the string is not coupled to the vector field. While it may still carry a chiral current (characterized by $Y$), it emits no electromagnetic radiation. Consequently, there is no energy injection into the IG, and the $21$-cm constraint vanishes entirely. The GW spectrum in this limit is determined solely by the chiral current effects.
\section{Conclusion}
\label{sec:conclusion}
In this work, we have explored the combined constraints on superconducting cosmic strings (SCSs) from gravitational wave (GW) observations and the Dark Ages global $21$-cm signal. Using the Charge-Velocity-dependent One-Scale (CVOS) model, we computed the stochastic gravitational wave background (SGWB) emitted by current-carrying loops, accounting for the suppression of GW emission due to vector radiation characterized by the dimensionless coupling $\tilde{e}$ and the current amplitude $Y$. We demonstrated that while large values of $\tilde{e}$ suppress the SGWB, allowing string tensions as high as $G\mu\simeq 10^{-8}$ to evade pulsar timing array (PTA) bounds such as those from NANOGrav, this parameter window is firmly constrained by an independent and astrophysically-clean probe: the Dark Ages $21$-cm signal. During the Dark Ages ($40\lesssim z \lesssim 200$), the intergalactic medium is free from astrophysical uncertainties, making it an ideal laboratory for detecting energy injection from exotic sources.

We recalculate the recently derived $21$-cm bounds~\cite{Si:2025vsj} onto the generic 
($\tilde{e}, Y, G\mu$) parameter space. Our analysis reveals that for $Y\simeq 0.67$ and $G\mu\simeq10^{-13}$, any coupling $\tilde{e}\gtrsim 10^{-7}$ produces sufficient ionizing radiation to completely erase the $21$-cm absorption signal at $z\sim 89$. For larger current amplitudes ($Y\simeq 1- 10^{-5}$), the constraints become significantly more stringent, restricting the allowed parameter space to $\tilde{e}\gtrsim 8\times 10^{-10}$. These bounds are orders of magnitude smaller than the coupling $\tilde{e}\sim 10^{-4}$ required to evade NANOGrav constraints~\cite{NANOGrav:2023hvm}, effectively closing the viable parameter window for SCSs. Our results highlight the complementarity between future lunar-based $21$-cm experiments, such
as FARSIDE~\cite{burns2019farsidelowradiofrequency}, DAPPER~\cite{Burns:2021ndk}, FarView~\cite{Burns:2021ndk}, and LuSee-Night~\cite{bale2023luseenightlunarsurface, 10906958}, and GW observatories. If future $21$-cm observations detect the standard Dark Ages absorption signal, they will place definitive upper bounds on $\tilde{e}$ across a wide range of $G\mu$ and $Y$, ruling out large portions of the SCS parameter space. Conversely, a suppressed or absent $21$-cm signal could provide compelling evidence for the existence of superconducting cosmic strings or other exotic energy sources.

\section*{Acknowledgments}
The author thanks Ivan Rybak for a useful discussion regarding the calculations of the power spectrum of superconducting cosmic strings. I also gratefully acknowledge Qaisar Shafi, Rishav Roshan, and Ahmad Moursy for engaging discussions that helped bring this project to fruition. The calculations were carried out using the heterogeneous computing platform HybriLIT (LIT, JINR).

\bibliographystyle{apsrev4-1}
\bibliography{refs}
\end{document}